# Spectroscopic evidence for polarons in partially occupied spin-orbit states of $YBa_2Cu_3O_7$

J.V. Acrivos[(1,2)] L. Nguyen[(2)], H.S. Sahibudeen[(2)], P. Nachimuthu[(3)] and M.A. Navacerrada[(2,4)]

[(1)]jacrivos@athens.sjsu.edu, [(2)]*S JSU, CA 95192,* [(3)]*LBNL-ALS and U. Nevada, NV89154,* [(4)]*U. Politecnica Madrid 28040 Spain*

The spin-orbit split white line, synchrotron X-ray spectra at the respective Cu and Ba $L_{2,3}$ and Ba $M_{4,5}$ edges in $YBa_2Cu_3O_7$ crystal, powders, films and derivatives are compared to those for standard $BaBr_2$ and CuO powders. The white line absorption integrated intensity, I ratios: $r_{3,2} = I(L_3)/I(L_2)$ and $r_{5,4} = I(M_5)/I(M_4)$ for $BaBr_2$ are equal to the ratio of core states degeneracy, but $r_{3,2}$(Ba in YBCO $_{powder}$) < 2 < $r_{3,2}$(Cu in CuO and YBCO $_{film}$) and $r_{5,4}$(Ba in YBCO $_{film}$) < 1.5 (Table I) indicate an apparent overpopulation of the Cu:$3d_{3/2}$ final states relative to the $3d_{5/2}$, and the Ba:$5d_{5/2}$ over the $5d_{3/2}$, and the $4f_{7/2}$ over the $4f_{5/2}$ at room temperature. The difference spectrum DEL=A(YBCO)-A(CuO) with a leading sharp peak, (HWHH of 0.2 eV) at $E_0$+0.5 ±0.1eV and the temperature changes in the atomic X-ray absorption fine structure A-XAFS observed at 0.08 to 0.07 nm near the heaviest atom, Ba suggest that charge polarization contributes to the final state band occupancy in YBCO.

*To whom all correspondence should be sent: jacrivos@athens.sjsu.edu, TEL 408 924 4972, FAX 408 924 4945

## INTRODUCTION

The synchrotron X-ray absorption spectra (XAS) of layered cuprates, where superconducting planes are intercalated between ionic and perhaps magnetic layers in $YBa_2Cu_3O_7$ and its derivatives (YBCO) are compared to the CuO and $BaBr_2$ powder spectra at the Cu and Ba $L_{2,3}$ edges and the Ba $M_{4,5}$ edges for the purpose to ascertain evidence of charge polarization in YBCO.

## EXPERIMENTAL

The samples are single crystals/powders grown at the Cavendish Laboratory[1] and 50 nm films that have mixed ab axes, grown epitaxially by sputtering in an oxygen atmosphere onto $SrTiO_3$, single crystal, (SC) and, bi-crystals (BC) with a 24 DEG ab grain boundary at the Complutense[2]. XAS in transmittance ($I_T$), in fluorescence (F) and total electron yield (TEY) were measured versus energy, E calibrated at SSRL station 2-3 relative to Cu film Cu:K-edge[3-5], and at the LBNL-ALS 6.3.1 Nachimuthu chamber relative to CuO. Cu:$L_3$ edge[6]. XAS, A = $\ln(I_0/I_T)$, $(F/I_0)$, $(TEY/I_0)$ when $I_0$ is the incident intensity and white line (WL) spin-orbit split transitions and WL integrated intensities, I are reported (Table I, FIG. 1 to 4).

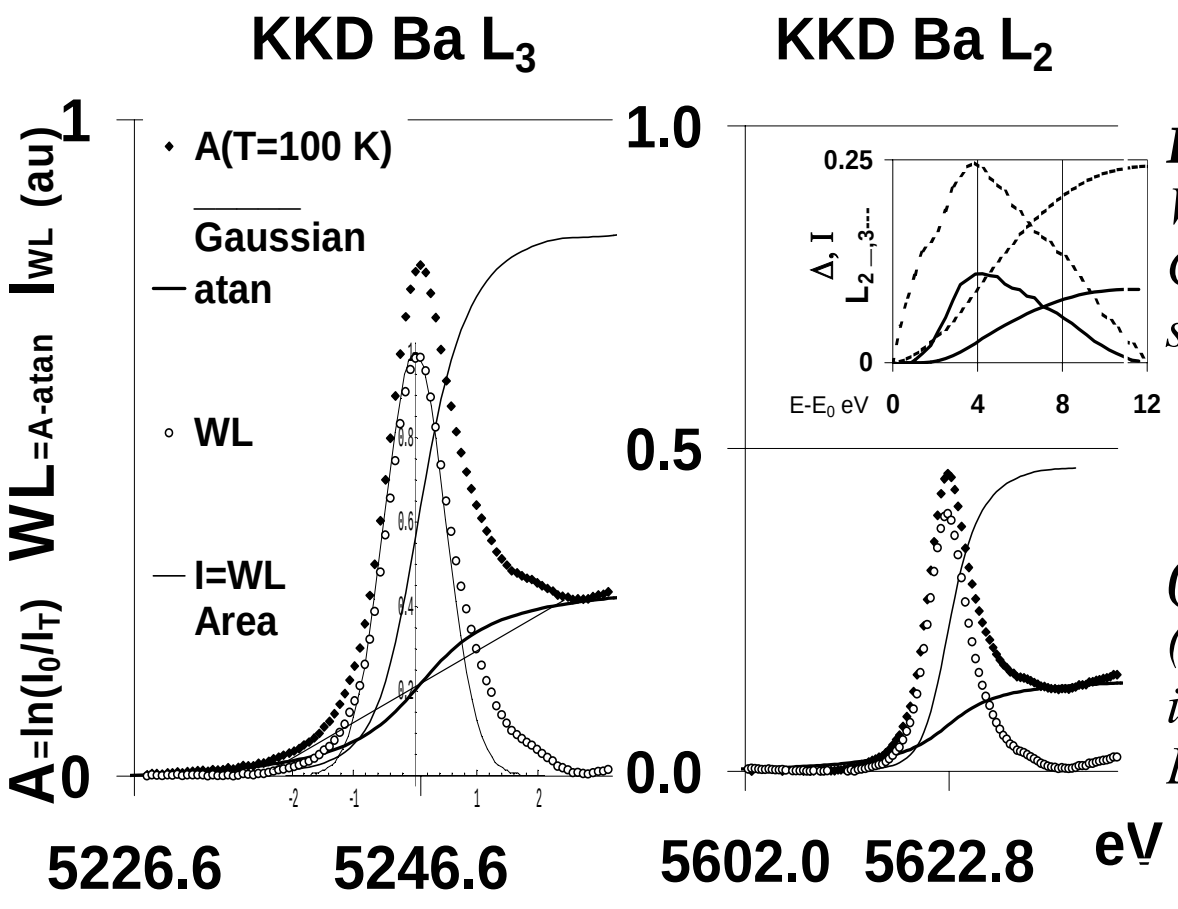

***FIG. 1*** *Ba $L_{2,3}$: A, WL =A–$A_{core\ to\ cont.}$ Gaussian, G shape fit to WL(E<$E_0$), Δ = WL-G, I*

***(a)*** *$KKD_{powder}$: ($Nd_{1.1}Ba_{1.9}Cu_3O_7$) insert shows Δ at $L_2$ and $L_3$ edges.*

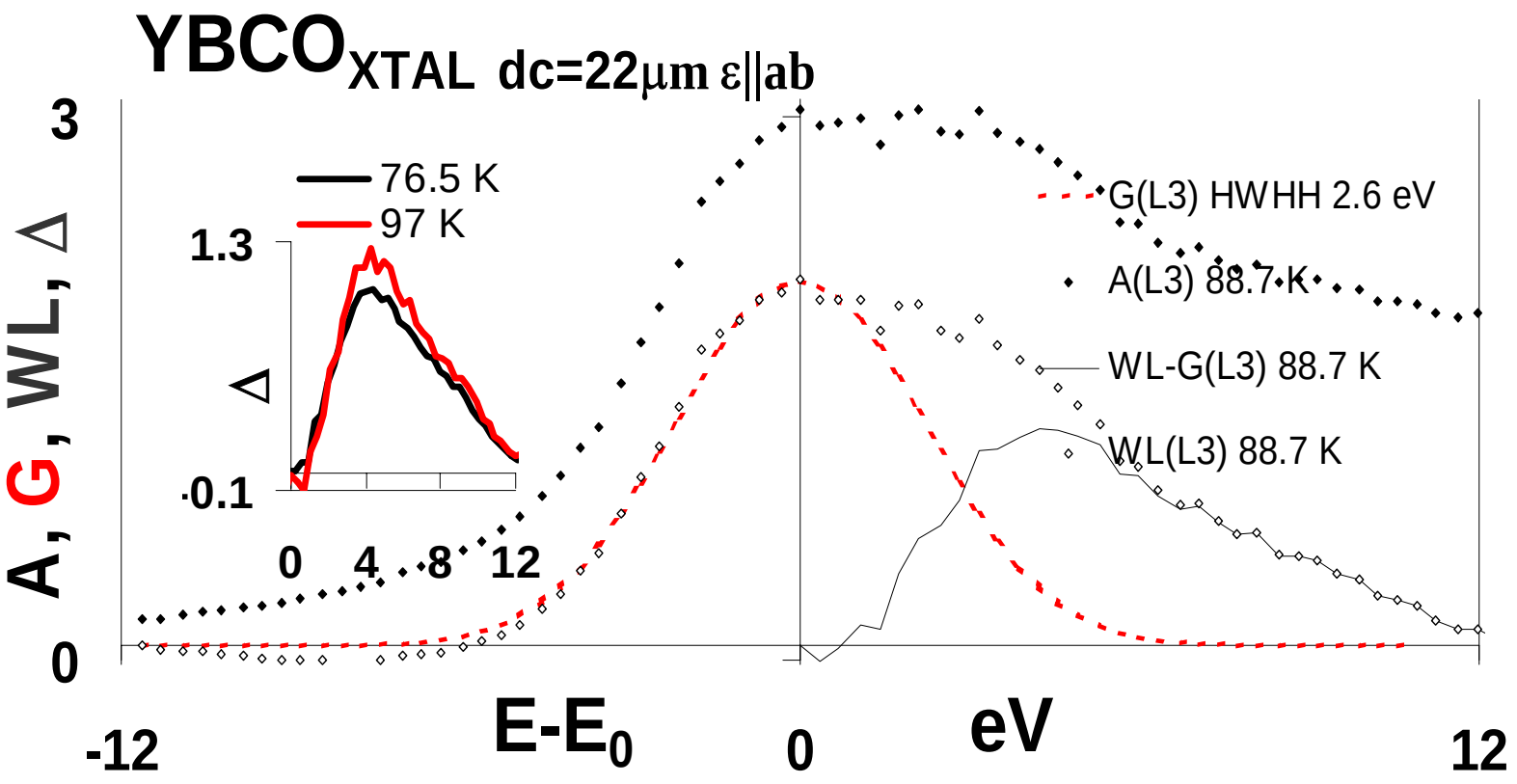

***FIG. 1 (b)*** *YBCO single crystal where insert shows structure in Δ decreasing below $T_c$.*

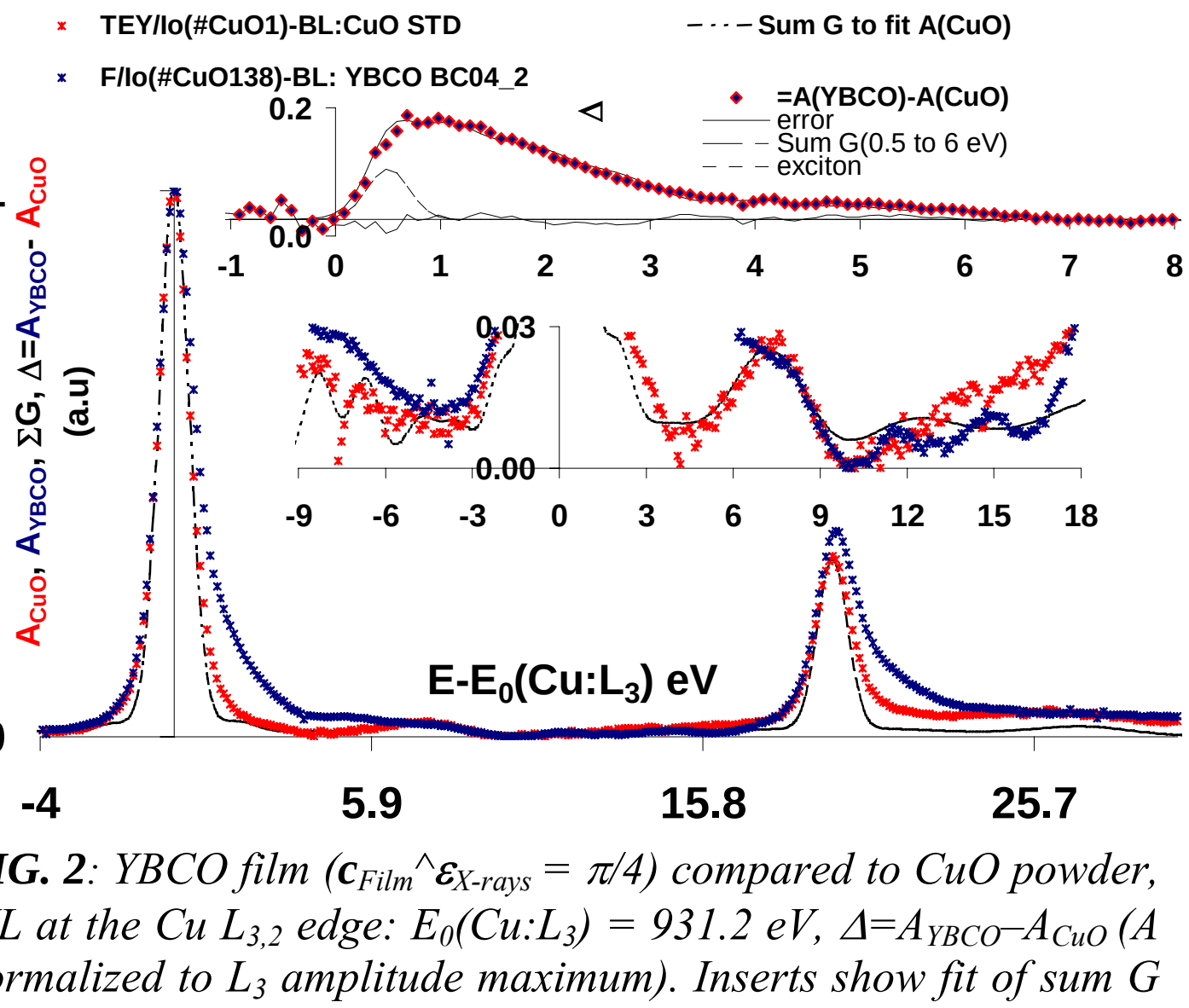

***FIG. 2****: YBCO film ($c_{Film}$^$\varepsilon_{X\text{-}rays}$ = π/4) compared to CuO powder, WL at the Cu $L_{3,2}$ edge: $E_0$(Cu:$L_3$) = 931.2 eV, Δ=$A_{YBCO}$–$A_{CuO}$ (A normalized to $L_3$ amplitude maximum). Inserts show fit of sum G shape peaks (HWHH~0.45eV) to $A_{CuO}$ with less than 2% error and to Δ with less than 1% error if an exciton peak is included.*

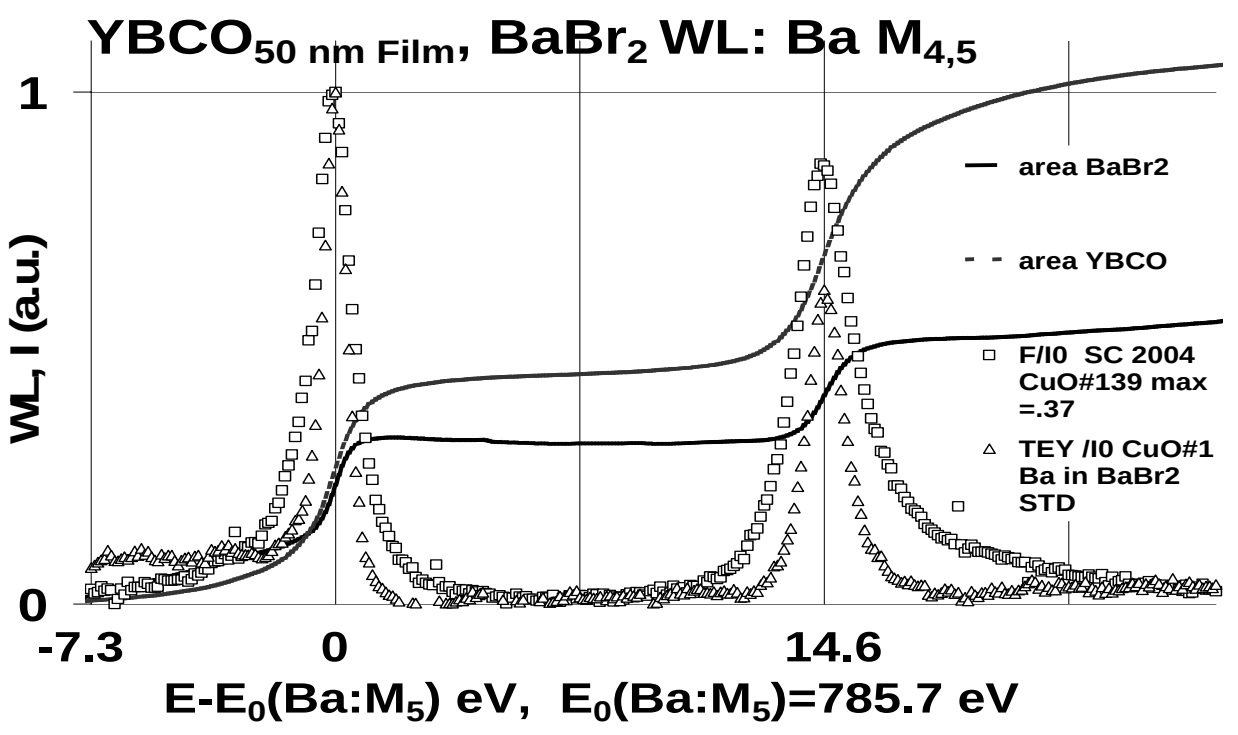

***FIG. 3**: WL at the Ba:$M_{4,5}$ edges for YBCO film ($\mathbf{c}_{Film}$^$\varepsilon_{X\text{-}rays} = \pi/4$) and $BaBr_2$ powder. Weak shoulders in $A_{BaBr2}$ are due to Br($L_2$)~1.599, Br($L_3$)~1.553 keV absorption from beam 2nd order harmonic. 3rd order harmonic ( Nb($L_3$) ~ 2.3705 keV), is absent.*

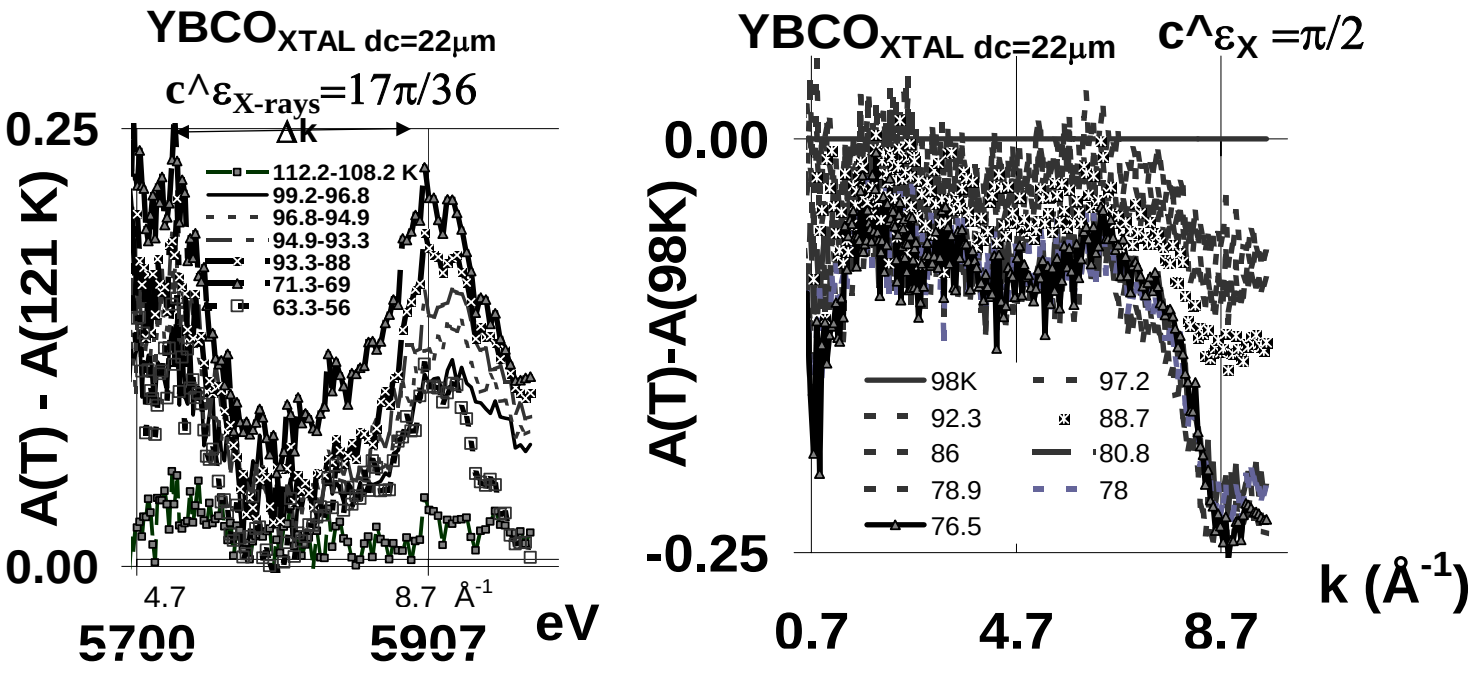

*(a)* *(b)*

***FIG. 4:** Change in atomic XAFS versus T, XTDAFS for $YBa_2Cu_3O_{7-\delta}$ single crystal at Ba $L_2$ edge through $T_c$: The XTDAFS measure the changes in XAFS scattering amplitude [ref.4, 5, 11]: $\chi(k)=-\sum_j N_j \sin(2\mathbf{k}.\mathbf{R_j}+\varphi(k))F(k,R_j)/kR_j^2$ caused by changes in the number density $N_j$ at $R_j$ with scattering amplitude F. When inter atomic distances do not change across $T_c$ their contributions to $\chi(k)$ cancel out in the XTDAFS [ref.4, 5], and the appearance of a short wave oscillation, $\Delta k$~4$Å^{-1}$ indicates $N_{j=p}$ T dependence at $R_j = r_p$ obtained from the period $2r_p\Delta k = 2\pi$ or $r_p \sim \pi/\Delta k$ ~ 0.07 to 0.08 nm (when $\varphi(k)$~constant in the interval $\Delta k$). **(a)** Increase in $N_p$ across $T_c$ when $\varepsilon_{X\text{-}rays}$ is in ab plane. **(b)** Decrease in $N_p$ across $T_c$ when $\varepsilon_{X\text{-}rays}$ is out of ab plane.*

Mass attenuation due to sample absorption at the two spin orbit split WL, determined by atomic cross sections[7] differs by less than 8 % in F, and is negligible in TEY and E > $10^3$ eV. The WL spectra are obtained by subtracting from A the background baseline (BL) and the extended X-ray absorption to the empty continuum states, X-ray absorption fine structure, XAFS[4-5] region (FIG. 1), except where the XAFS contribution from the previous edge is less than 2% at the Cu:$L_2$ and Ba $M_4$ edges in both YBCO and CuO (FIG. 2, 3).

## RESULTS/DISCUSSION

The normalized A difference spectrum $\Delta = A_{YBCO} - A_{CuO}$ gives information of interdependent effects on the states near the Fermi energy in YBCO, e.g., crystal field symmetry different from CuO[8], polarization effects[9], and electron density in h

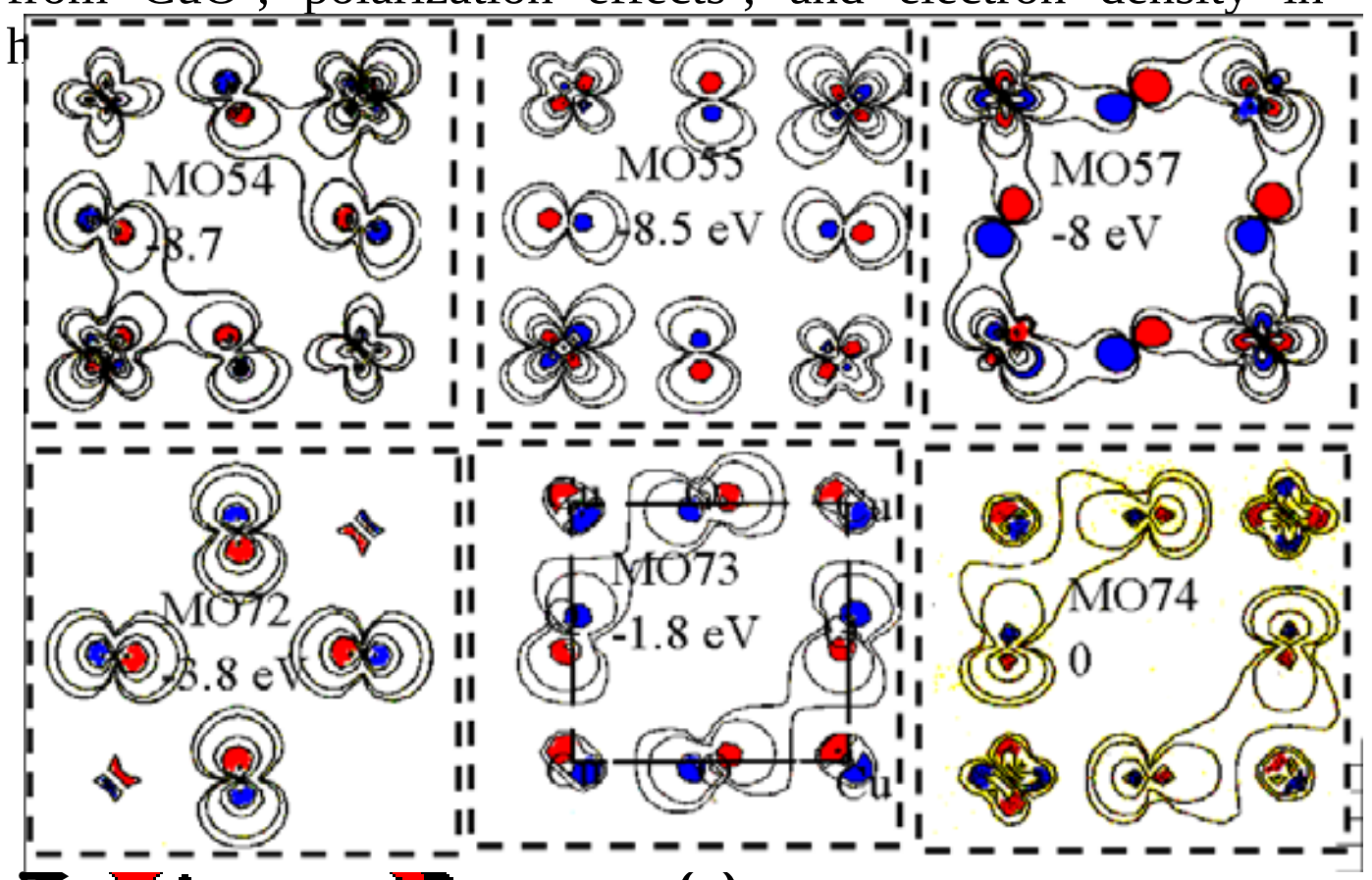

(a)

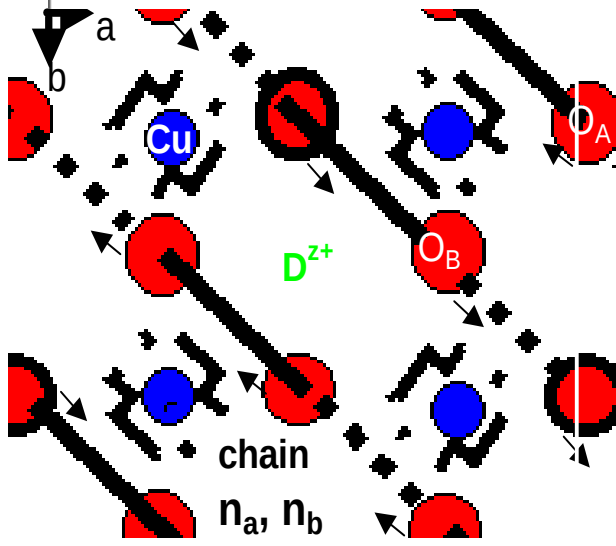

(b)

***FIG. 5:** Highest occupied MO showing the electron density, $\rho_e$ continuous along O:$2p_{a\pm b}$ diagonals, but discrete at Cu atom that can contribute to both polarization and periodic lattice distortions (PLD), and absorption through shake-up at ~$E_0$+2 to ~ $E_0$+8eV at the Cu, Ba $L_{2,3}$ edges:*

***(a)** Contours of $\rho_e \geq 10^{-3}b^{-3}$, $\varepsilon-\varepsilon_{HOMO}$ = 0 to – 8.7 eV for $Cu_4O_4$ lamella in $(T'Nd_2CuO_4)_{18}$ [full Hartree-Fock, SCF calculation ref. 10]. **(b)** Schematic of highest occupied MO with O:$2p_{a\pm b}$ continuous overlap. The arrows indicate direction of PLD along a±b diagonals observed in XRD [ref. 3]. MO phase > 0: ▬, and < 0: ■■. $D^{z+}$ represent nominal $Y^{+3}$ or $Ba^{+2}$ with Cu apical O in Ba layer*

(i) Evidence of charge polarization is obtained from atomic A-XAFS[11] at distances less than an Å as follows: When the inter-atomic distances remain constant versus T, the changes in electron density, $\rho_e$ near the heaviest element in YBCO, Ba are detected by the XAS temperature difference spectra, XTDAFS[4,5] amplitude versus k(bohr$^{-1}$)=(2(E-$E_0$) H)$^{½}$ (FIG. 4). The observed oscillations identify a change in $\rho_e$ near $r_p$ ~ $\pi/\Delta k$ ~ 0.07 to 0.08 nm from the Ba atom, across the transition to superconductivity, $T_c$ that reverses sign when the polarized synchrotron electric field, $\varepsilon_{X\text{-}rays}$ is rotated out of the crystal ab plane. This indicates that polarons [9,12] of radius $r_p$ are involved.

(ii) The formation of polarons determines the symmetry and final states occupation and consequently the WL spectra relative intensity as follows: The XAS white lines (WL) observed at the Cu:$L_{2,3}$ edges in CuO and cuprate superconductors show structure consisting of a main peak with gaussian, G shape (HWHH~0.45eV) plus weaker ones. The $A_{CuO}$ additional structure indicates that the crystal field shifts in Cu:$3d^9$ may be described by G shape peaks with 3% of $A_{CuO,max}$ and HWHH~0.45eV at $E_0$±1 to ±9eV, (FIG.2 insert) expected to be more complex than the $V^{+4}$:$3d^1$ cubic crystal field splitting[8c]. The crystal orientation in $\varepsilon_{X\text{-}rays}$ determines the structure of spin-orbit WL pairs[3-6,8], but their separation, $\Delta E(Cu:L_{2,3})$ is constant and dominated by the atomic core splitting[8a]: The WL separation, $\Delta E(Cu:L_{2,3})$ = 19.7 eV (FIG. 2) is the same as in other Cu compounds[6] though the crystal field induced structure depends on the $3d^N$ filled Cu oxidation states[8c]. Crystal field split peaks appear from $E_0$+0 to $E_0$+8eV in Cu metal, partially oxidized Cu film and Cu quantum dots respectively[6], but the 10% asbsorbance present at $E_0$+7.2eV for both YBCO and CuO suggests that they have the same oxidation state $Cu^{+2}:3d^9$ in a distorted lattice. Their difference spectrum $\Delta=A_{YBCO}-A_{CuO}$ vanishes to within less than 1% for $E<E_0$ at the $Cu(L_{2,3})$ edges but differs by as much as 20% $WL_{max}$ for $E>E_0$ (insert FIG. 2) and may be described by a sharp (HWHH~0.2eV) G shaped leading peak at $E_0+0.5\pm0.1$eV distinct from broader ones (HWHH~0.45eV) at $E_0$ ±1 to ±7eV (FIG. 2 insert). The first is too sharp to be assigned to a crystal field peak, the broader ones may be due to $Cu^{+3}:3d^8$ crystal field shift lines if the oxidation state is present in YBCO, and/or XAS shake-up processes from $CuO_2$ layer MO orbitals below the edge by that amount (FIG. 5)[10]. The $Cu^{+3}:3d^8$ should be similar to $Ni^{2+}$ WL spectra at the $L_{2,3}$ edges where all the cubic crystal field split lines show comparable amplitudes[8c] while the $Cu^{+3}$ splitting is reported to be twice as large[8b] as for $Ni^{2+}$. The source of the amplitude $\Delta$ appears to be due to charge polarization in the highest filled MO (FIG. 5) that in turn governs the crystal field symmetry.

At the Ba edges the WL can be fitted to G shapes (HWHH~2.5eV) for $E<E_0$, with $\Delta E(Ba:L_{2,3})$ = 376.2eV and $\Delta E(Ba:M_{4,5})$ = 14.0eV for YBCO and $BaBr_2$ (FIG. 1, 3). The difference spectrum $\Delta=WL_{YBCO}$-G for $E>E_0$ is orientation dependent with amplitudes 30% $WL_{max}$ for $\varepsilon_{X\text{-rays}}$ in the ab plane that reduces to less than 10% of $WL_{max}$ when $\varepsilon_{X\text{-rays}}$ is rotated out of the ab plane by $\pm\pi/18$. For $\varepsilon_{X\text{-rays}}$ in the ab plane $\Delta$ shows a structure with a sharp leading peak similar to that observed at the $Cu:L_3$ edge (Fig. 1b).

The 2p effective atomic number is obtained in the hydrogen-like approximation[8a] from $\Delta E$ (Table I) where the Pauling rules indicate that the Cu:2p electron nuclear charge is shielded by ~ $1s^22s$ shells, the Ba:2p by ~$1s^2$ shells and the Ba:3d by ~$1s^22s^22p^63s$ shells.

(iii) The final band state population, the crystal field and orientation in $\varepsilon_{X\text{-rays}}$ determines the WL transition probability[8]. All orientations in $\varepsilon_{X\text{-rays}}$ are sampled in powders, and when all the accessible final states are empty in the ionic reference $BaBr_2$, the ratio of spin-orbit split WL, integrated intensity is proportional to the core states degeneracy: $r_{3,2} = I(L_3)/I(L_2) = 2$ and $r_{5,4} = I(M_5)/I(M_4) = 3/2$ (Fig. 3 and Table I). At the Cu $L_{2,3}$ edges: $r_{3,2}$(Cu:$3d^9$ in CuO powder) = 2.7 for the Gaussian shaped WL (Table I, Fig. 2) is qualitatively explained by the presence of only one empty $3d_{5/2}$ level at the top of the conduction band available for $L_3$ transitions but, only $4d_{3/2}$ final states are available for $L_2$ transitions. The YBCO $Cu:L_{2,3}$ WL relative intensities (Table I) are similar to those for CuO, in agreement with a $Cu:3d^9$ oxidation state. At the $Ba:L_{2,3}$ edges the XAFS amplitudes, which measure the transition probability from the same core states to the empty continuum and obey the same orientation dependence as the WL, obtain for all samples a relative intensity proportional to core state degeneracy: $r_{3,2}$(Ba XAFS for YBCO) = 2 to within 4% accuracy (Table I). The significant deviations in YBCO WL: $r_{3,2}$(Ba in YBCO KKD powder) < $r_{3,2}$(Ba in YBCO single crystal) < 2 and $r_{5,4}$(Ba in YBCO film) < 1.5 (Table I) are explained by different transition probability due to final state $5d^1$ and $4f^1$ apparent occupation, respectively due to polarization. Since all orientations in $\varepsilon_{X\text{-rays}}$ are sampled for $r_{3,2}$(Ba in YBCO KKD powder), an apparent overpopulation of the $5d_{5/2}$ states relative to the $5d_{3/2}$ final states must be caused by polarons at $r_p$, and the same is true for the single crystal, though crystal field shifts and mixed valence effects are not resolved for a WL with HWWH ~2.5eV. Similarly $r_{5,4}$(Ba in YBCO film) < 1.5 indicates the effect of orientation, an apparent over population of $4f_{7/2}$ relative to the $4f_{5/2}$ levels and/or mixed valence.

The apparent difference in occupation of the spin orbit split final state bands in the non-magnetic YBCO[12b] may be qualitatively explained by the formation of $e_1e_2^=$ pairs with $j = j_1+j_2 = 0$ in different layers where O states are involved in the highest occupied MO (FIG. 5)[15].

(iv) Line shape analysis gives additional evidence of charge polarization as follows: The difference spectrum, $\Delta(Cu:L_{3,2})$ leading sharp peak (HWHH~0.2 eV) at $E_0+0.5\pm0.1$ eV (first insert FIG. 2) suggests that in YBCO 50 nm film the ejected electron and X-ray hole form pairs, via a hydrogen like potential, in atomic units[9a]:

$$V(r) = -1/(K\,r)\ \text{H} \quad \text{when } r > r_p$$
$$= -1/(K\,r_p)\ \text{H} \quad \text{when } r < r_p \qquad (1)$$

where $r_p$ is the radius of polarized charge and $1/K = 1/\varepsilon_\infty - 1/\varepsilon_0$, when $\varepsilon_\infty$ and $\varepsilon_0$ are the material high frequency and static dielectric constants. In transition metal oxides $\varepsilon_0$~10, and as high as 30 in $La_2CuO_4$[9c] is meaningless in the region between atoms and $\varepsilon_\infty$~5 to 8[16]. The hydrogen-like energy level eigenvalues[8a] in (1): $E_n = -1/(2n^2K^2)$, n = 1 to ∞ give rise to transitions at hν where:

$$h(\nu-\nu_0) = E_n - E_1 = \tfrac{1}{2}(1-1/n^2)/K^2\ \text{H},\ n \geq 2. \qquad (2)$$

An X-ray exciton head at:

$$h(\nu-\nu_0)=(0.5\pm0.1\text{eV})/(27.21\text{eV/H})=3/8/K^2\ \text{H} \qquad (3)$$

obtains $K = 4.2 \leq \varepsilon_\infty$ and $E_n = -0.029/n^2$H predicts a sharp Rydberg series that ends sharply at $E_0$+0.8eV. The average separation $1/\langle r^{-1}\rangle_{n=1} = K$*0.053nm/bohr = 0.22nm > CuO bond distance 0.195nm, when added to $r_p$~0.07 to 0.08nm near Ba indicates an effect on their outer electron shell population is a possibility. This is similar to the formation of excitons in layer dichalcogenides observed in i.r. spectra[9, 12a].

(v) Many electron interactions in bulk metals that are responsible for enhanced absorption to levels close to the Fermi energy[17] a few meV above $E_0$ are not be resolved here.

## CONCLUSION

Spin-orbit split white lines in YBCO indicate an apparent overpopulation of the $Cu:3d_{3/2}$ final states relative to the $3d_{5/2}$ as in CuO, and of the $Ba:5d_{5/2}$ final states relative to the $5d_{3/2}$ as expected to $5d^1$ occupation. The formation of X-ray excitons, observed at the Cu $L_{2,3}$ edges, and the orientation dependent changes in atomic XAFS at the Ba $L_2$ edge of a YBCO single crystal suggest that the formation of polarons is responsible for the apparent population of states near the Fermi level and these change near $T_c$.

**TABLE I:** Core $Z_{eff}$ and relative intensity of spin orbit split WL, WL($E<E_0$) and XAFS[a,b]

| WHITE \ LINE | Changes in Occupied Shells | Changes in Occupied Shells | $Z_{2p,eff}$ $\Delta E_{experimental}$ (H) | $Z_{3d,eff}$ [a] | $I(L_3)$[b] / $I(L_2)$ $\pm 0.1$ | | | $I(M_5)$ / $I(M_4)$ |
|---|---|---|---|---|---|---|---|---|
| Edge: | $L_3$ | $L_2$ | | | **WL** | **XAFS** | **GAUSSIAN** | **WL** |
| ELEMENT (Z) | | | | | **$E_0$>E>$E_0$** | | **E<$E_0$** | |
| CORE SHELL | $.(2p_{3/2})^4.$ | $.(2p_{1/2})^2.$ | | | | | | |
| FINAL: Cu (29) | $.(2p_{3/2})^3.(nd_{5/2})$ $n\geq 3$ | $.(2p_{1/2}).(nd_{3/2})$ $n\geq 4$ | **25.6** | | | | | |
| **Sample Alignment: c^$\varepsilon_{X\text{-}rays}$** | | | 0.726 | | | | | |
| CuO powder, TEY/$I_0$ | | | | | **2.7** | | **2.7** | |
| YBCO Film F/$I_0$: $\pi/4$ | | | | | **2.8** | | **2.6** | |
| YBCO Film TEY/$I_0$: $\pi/2$ | | | | | **3** | | **2.8** | |
| FINAL: Ba (56) | $.(2p_{3/2})^3.(nd_{5/2})$ $n\geq 5$ | $.(2p_{1/2}).(nd_{3/2})$ $n\geq 5$ | **53.6** | | | | | |
| **Sample Alignment: c^$\varepsilon_{X\text{-}rays}$** | | | 13.82 | | | | | |
| KKD YBCO powder, $I_0/I_T$ | | | | | **1.7** | **2** | **1.5** | |
| YBCO Single Crystal, $I_0/I_T$: $17\pi/36$ | | | | | **1.7** | **2** | **1.5** | |
| Edge: | $M_5$ | $M_4$ | | | | | | |
| CORE SHELL | $.(3d_{5/2})^6.$ | $.(3d_{3/2})^4.$ | | | | | | |
| FINAL: Ba (56) | $.(3d_{5/2})^5.$ $(nf_{7/2})$ $n\geq 4$ | $.(3d_{3/2})^3.$ $(nf_{5/2})$ $n\geq 4$ | | **45.1** | | | | |
| **Sample Alignment: c^$\varepsilon_{X\text{-}rays}$** | | | | 0.544 | | | | |
| $BaBr_2$ powder, TEY/$I_0$ | | | | | | | | **1.6** |
| YBCO Film F/$I_0$ : $\pi/4$ | | | | | | | | **1.1** |

[a] *The $\Delta E_{experimental}$ in atomic units, are set equal to the spin-orbit energy difference [ref. 8a] :*
$\Delta E_{SO}(L_{2,3}) = [-Z^4_{nd,eff}/715 + Z^4_{2p,eff}/32]/c^2$ H, $\Delta E_{SO}(M_{4,5}) = [-Z^4_{nf,eff}/2150.4 + Z^4_{3d,eff}/405]/c^2$ H.

[b] *The finite sample thickness, t absorption: $abs(E,t) = \int_0^t e^{-\eta(E)t'}dt'/\int_0^t dt' = [1- e^{-\eta(E)t}]/\eta(E)t$, when $\eta(E) = \Sigma_{i=components}\, x_i*\alpha_i(E_{edge})*$sample density, $x_i$ = element i weight fraction, $\alpha_i(E)$ = component i cross section [ref. 7], obtains a ratio: $abs(E_{l+1/2})/abs(E_{l-1/2}) \sim 1.1$ at both the Cu $L_{2,3}$ and Ba $M_{4,5}$ edges in F, approaches 1 at the Ba $L_{2,3}$, and TEY surface probe, and is within the area measurements uncertainty limits. The XAFS amplitudes extrapolated to $E_0$, $\mu_0$ obtain intensity ratios equal to the free$Ba^{2+}$ ion values to within $\pm 4\%$ error.*

## ACKNOWLEDGEMENT

Work was supported by NSF, NATO, Dreyfus Foundation at SJSU and DOE at SSRL and LBNL-ALS. Thanks are due to Dr. K.K. Singh and Professor W.Y. Liang for samples and encouragement, and to Prof. A.S. Alexandrov and Dr. U. Bergmann for their respective knowledge in superconductivity[9] and $Ni^{2+}$:$3d^8 \sim E_0$+1eV crystal field shifts[18].

## REFERENCES

[1]C.T. Lin, W.Zhou and Y. W. Liang, *Physica C*, **195** 291 (1992)

[2] M. A. Navacerrada, M. L. Lucía and F. Sánchez-Quesada, *Europhys. Lett* **54**, 387 (2001)

[3] (a) M.A. Navacerrada and J.V. Acrivos, *NanoTech 2003*, **1**, 751 (2003), (b) H.S. Sahibudeen, M.A. Navacerrada and J.V. Acrivos, *NanoTech* **3**, in press (2005)

[4] J.V. Acrivos, *Solid State Sciences*, **2**, 807-820 (2000)

[5] J.V. Acrivos et al., *Microchemical Journal,* **71**, 117-131 (2002)

[6] P.Nachimuthu et al., *Chem. Mater.* **15,** 3939 (2003)

[7] Center for X-Ray Optics, http://www.cxro.lbl.gov

[8](a)L.I. Sciff, "Quantum Mechanics", McGraw Hill Book Co. Inc, 1949, p 280;(b) A. Abragam and B.Bleaney, *"Electron Paramagnetic Resonance of Transition Ions"*, Oxford (1970); (c) F.M.F. de Groot, J.C. Guggle, B.T. Thole and A.G. Sawartzky, *Phys. Rev* **B42,** 5459 (1990)

[9] (a) N.F. Mott, *"Metal Insulator Transitions"*, Taylor and Francis, ltd, 1974, 1990; (b) A.S. Alexandrov and N.F. Mott, *"Polarons and Bipolarons"*, World Scientific, Singapore (1995); (c) private communication (2005)

[10] J.V. Acrivos and O. Stradella, *International Journal of Quantum Chemistry*, **46**, 55(1993)

[11] (a) A.L. Ankudinov and J.J. Rehr, *J. Phys. IV France 7*, **C2**- 121 (1997); (b) A. Kodre, I. Arçon and R. Frahm, *ibid.* **C2**-195 (1997); (c) Y.A. Babanov, A.V. Ryazhkin and A.F. Sidorenko, *ibid*, **C2**-277 (1997)

[12] (a) J.A. Wilson and A.D. Yoffe, Adv. In Physics **18,** 193 (1969); (b) T.J. Smith et al., *J. Magnetism Magn. Mater.*, **177**, 543(1998)

[13] Z.-X. Shen, W.E. Spicer et al., *Science* **267**, 343 (1995)

[14] (a) A. Bianconi et al, *Phys. Rev.* **B38,** 7196 (1988); (b) N. Nucker et al., *Phys. Rev.* **B51**, 8259 (1995)

[15] J.V. Acrivos, *Microchemical Journal,* in press (2005)

[16] G. Koster, T.H. Geballe and B.Moyzhes, *Phys. Rev.* **B66**, 085109 (2004)

[17] M. Hentschel, D. Ullmo and H. Baranger, *Phys. Rev. Lett.* **93**, 176807 (2004)

[18] P. Glatzel, U. Bergmann et al., *J. Am. Chem. Soc.*, **124**, 9668 (2002)